\documentstyle[twoside,fleqn,espcrc2,epsfig]{article}

\title{Finite-size scaling study of the $d=4$ site-diluted Ising 
        model\thanks{Partially supported by CICyT (AEN94-0218,
        AEN96-1634). JJRL is granted by  EC HMC (ERBFMBICT950429).
        Presented by HGB.}.}

\author{H.~G.~Ballesteros\address{Dep. de F\'{\i}sica Te\'orica I, 
        Facultad de CC. F\'{\i}sicas, 
        U. Complutense de Madrid, 28040 Madrid, Spain.},
        \setcounter{address}{0} 
        L. A. Fern\'andez\addressmark, 
        \setcounter{address}{0} 
        V.~Mart\'{\i}n-Mayor\addressmark, 
        \setcounter{address}{0} 
        A.~Mu\~noz Sudupe\addressmark,        
        G.~Parisi\address{Dip. di Fisica and INFN, 
        U. di Roma ``La Sapienza'', P.~A.~Moro  2, 
        00185 Roma, Italy.} and 
        \setcounter{address}{1} 
        J.J.~Ruiz-Lorenzo.\addressmark}

\begin{document}

\begin{abstract}

We study the four dimensional site-diluted Ising model using 
finite-size scaling techniques. We explore the whole parameter space 
(density-coupling) in order to determine the Universality Class 
of the transition line. Our data are compatible with Mean Field behavior
plus logarithmic corrections.

\end{abstract}

\maketitle

\section{Introduction}

If a pure system has a specific heat exponent $\alpha>0$ the
Universality Class changes when the dilution is introduced in the
model (Harris criterion~\cite{HARRIS}), while it remains unchanged if
$\alpha<0$ (i.e. the Universality Class is that of the pure model).
In the Ising model in 4 (or 2) dimensions $\alpha=0$ and the Harris
criterion does not apply.

Perturbative renormalization group (PRG)
computations for $d=4$ predict Mean Field with Logarithmic Corrections.
On the other hand, previous Monte Carlo (MC) results  pointed to non Mean
Field behavior~\cite{PARU}. In $d=2$ there are also MC studies that
conclude a change of the Universality Class.

We describe here the results of a higher statistics MC
study~\cite{ISDIL4D}. A Finite-Size Scaling (FSS) approach has been used in
order to study large lattices in the critical region. Results on the
$d=2$ case are also briefly described~\cite{ISDIL2D}.

\section{The model}

We work in a hypercubic four dimensional lattice. The action is:
\begin{equation} 
S=-\beta\sum_{<i,j>} {\epsilon_i \epsilon_j} \sigma_i \sigma_j, 
\end{equation}
where $\epsilon_i$ are quenched uncorrelated random variables whose
value is 1 with probability $p$ and 0 otherwise.

For each $\{ \epsilon_i \}$ configuration (sample) we perform an Ising
model simulation.

There are two types of averaging. The first corresponds
to averaging in Ising configurations, and will be denoted with
brackets, the second is associated to the
$\epsilon_i$ variables (sample average) and will be denoted by overlines.
We first perform the Ising average, then the sample one.

\subsection{Observables}

For each spin configuration we measure the magnetization and first
neighbor energy, defined respectively as
\begin{equation} 
{\cal M}=\frac{1}{V}\sum_i \epsilon_i\sigma_i,\quad 
{\cal E} =\sum_{\langle i,j\rangle}\epsilon_i\sigma_i\epsilon_j\sigma_j.
\end{equation}

We have focused our study in the following mean values
(specific heat, susceptibility, Binder parameter, and correlation
length, respectively)
\begin{eqnarray}
C&=&V^{-1} \left( \overline{\langle{\cal E}^2 \rangle}  
        -\overline{\langle{\cal E} \rangle} ^2 \right)\ , \\
\chi&=&V\overline{\left\langle {\cal M}^2 \right\rangle}\ , \\
g_4&=&\frac{3}{2}-\frac{1}{2}\frac{\overline{\langle {\cal M}^4\rangle}}
           {\overline{\langle {\cal M}^2 \rangle}^2}\ , \\
\xi&=&\left(\frac{\chi/F-1}{4\sin^2(\pi/L)}\right)^{\frac{1}{2}}\ ,
\end{eqnarray}
where $F$ is the Fourier transform of the magnetization at $k=\frac{2\pi}{L}$.

\section{Finite-size scaling techniques}

In a finite lattice at the the critical region, the FSS 
ansatz states that
\begin{equation}
\langle O(L,\beta) \rangle=L^\frac{x_O}{\nu}
\left[F_O\left(\frac{\xi(L,\beta)}{L}\right) 
+O(L^{-\omega})\right],
\end{equation}
where
$\omega$ is the corrections-to-scaling exponent,
$F_O$ is a (smooth) scaling function and $x_O$ is the critical exponent.
For instance, $x_{\chi}={\gamma}$, $x_{\xi}={\nu}$, and 
$x_{\partial_{\beta}\xi}={\nu+1}$.

We study the quotient of $O(L_1)$ and $O(L_2)$
\begin{equation}
Q_O\equiv\frac{\langle O(L_2,\beta)\rangle}{\langle O(L_1,\beta)\rangle}=
s^\frac{x_O}{\nu}\frac{F_O(\frac{\xi(L_2,\beta)}{L_2})}
       {F_O(\frac{\xi(L_1,\beta)}{L_1})}+O(L^{-\omega}) 
\end{equation}
where $s=\frac{L_2}{L_1}$. 
The unknown 
$F_O$ can be  eliminated in this way: we look for 
$\beta({L_1,L_2})$ such that 
$\frac{\xi_{L_2}}{L_2}=\frac{\xi_{L_1}}{L_1}$ so
\begin{equation}
\left.Q_O\right|_{Q_{\xi}=s}=s^\frac{x_O}{\nu} + O(L^{-\omega}).
\end{equation}

Now, this picture is slightly modified due to the presence 
of logarithmic corrections.
Using the PRG analysis we obtain for the diluted model~\cite{ISDIL4D}
\begin{equation}
\begin{array}{ccl}
\xi&\propto&L(\log L)^{\frac{1}{8}} \quad \\
\partial_\beta \xi &\propto&L^3 \left(\log L \right)^{\frac{1}{4}}
\left(\frac{\xi}{L} \right)^3 
{\mathrm e}^{-2 \sqrt{\frac{3 \log L}{53}}} \raisebox{-2.5ex}\quad\\
\chi&\propto&L^2(\log L)^{\frac{1}{4}+\frac{1}{106}} \raisebox{-1.7ex}\quad \\
C&\propto&(\log L)^{\frac{1}{2}}\, {\mathrm e}^{-\sqrt{\frac{48}{53}|\log L|}}
\end{array}
\end{equation}

The scaling behavior for the pure model is:
\begin{equation}
\begin{array}{ccl}
\xi&\propto&L(\log L)^{\frac{1}{4}}\\
\chi&\propto&L^2(\log L)^{\frac{1}{2}}\\
C&\propto&(\log L)^{\frac{1}{3}}
\end{array}
\end{equation}

It can be shown that $\xi(L,\beta,p)/L$ remains as the scaling variable. 

The expected leading logarithmic correction for the $\nu$ exponent goes as
$1/\log L$ for the pure model, but changes to $1/\sqrt{\log L}$
for the diluted case. For the $\eta$ exponent, the correction is
always order $1/\log L$.

\section{Numerical methods}

We store individual measures of the energy and magnetization for
extrapolating in a neighborhood of the simulation parameters.

\begin{figure}[t]
\centering\epsfig{file=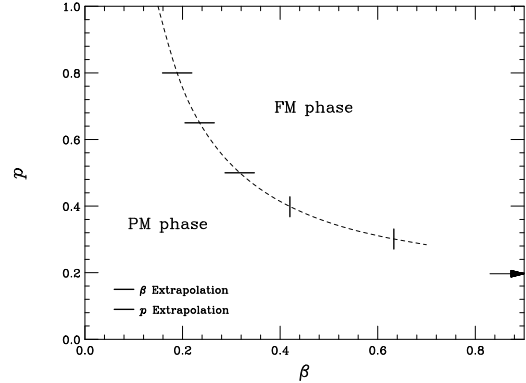,width=0.68\linewidth,angle=90}
\vglue -1cm
\caption{Phase diagram of the $d=4$ site-diluted 
Ising model. The tics are plotted at the simulated values along 
the extrapolation direction. The
black arrow indicates the percolation limit.}
\label{diagrama}
\end{figure}
The $\beta$-derivatives are obtained from
\begin{equation}
\partial_\beta \overline{\langle \cal O\rangle}=
\overline{\partial_\beta\langle \cal O\rangle}=
\overline{\left\langle_{\vphantom{|}} \cal OE \right\rangle}- \overline{\langle \cal O  \rangle}\overline{ \langle \cal E \rangle},
\end{equation}
which is a biased estimator. 
The statistical error behaves as $1/\sqrt{N_S}$, $N_S$ being the number
of samples and the bias as $1/N_I$ 
($N_I$ being the number of Ising independent measures in a sample). 
In our calculations $\sqrt{N_S} \sim N_I$  so a $N_I\to\infty$ 
extrapolation is performed.

As we gained statistics in a large number of samples with slightly
different number of filled sites, in addition to a $\beta$-extrapolation
a $p$ one is also possible.
 
The probability of finding $q$ site density for dilution $p$ is binomial.

The observable value from a set of $N_S$ samples at $p'$ near $p$  is
\begin{equation}
\begin{array}{l}
 \overline{\langle \cal O\rangle} (p',\beta)=\\
\quad\frac{1}{N_S}{\displaystyle\sum_i^{N_S} \left(\frac{p'}{p}\right)^{q_iV}
 \left(\frac{1-p'}{1-p}\right)^{(1-q_i)V} \langle {\cal O} \rangle_i(\beta)}
\end{array}
\end{equation}

It is also possible to compute $p$-derivatives but the statistical
error is much larger than for the $\beta$-derivatives (eight times
typically).

\section{Results}

We use cluster methods to update the spin variables.
The diluted model is simulated in lattice sizes $L\leq32$.
We also studied the pure one ($L\leq64$) as a check.  

We generate 10,000 samples at $p$=0.8, 0.65, 0.5, 0.4, 0.3.
For each sample we take 100 nearly independent measures after equilibration.

Let first assume hyperscaling (there are not logarithmic corrections).
We use $\partial_\beta \xi$ and $\chi$ to obtain the critical exponents.
In the $\nu$ case we perform a $L\to\infty$ extrapolation 
(see table \ref{NUNAIVE}), 
using $\omega=1$ (near the percolation value~\cite{PERC}).
For comparison, in the pure model we obtain  $\nu=0.5019(14)$.

For $\eta=2-\gamma/\nu$ we find a weaker evolution with $L$ and 
the dilution (see ref.\cite{ISDIL4D}).

We clearly discard the percolation values ($\nu=0.686(2)$ and
$\eta=-0.094(3)$ \cite{PERC}) and a new fixed point for all the critical
line is unlikely.

\begin{figure}[t]
\centering\epsfig{file=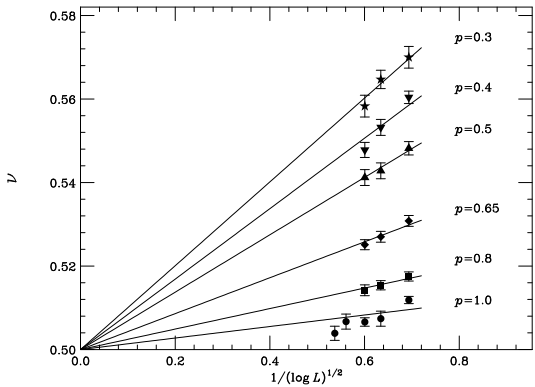,width=0.68\linewidth,angle=90}
\vglue -1 cm
\caption{Exponent $\nu$ estimation for $(L,2L)$ pairs.
The solid lines are linear fits constrained to be $\nu=0.5$ 
in the $L\to\infty$ limit.}
\label{NU1-logL}
\end{figure}

Weak universality ($\nu$ varying on the line) cannot be ruled out, but
a more economic picture is a MF behavior plus logarithmic corrections
(see fig. \ref{NU1-logL}).
\begin{table}[tbp]
\footnotesize
\begin{center}
\caption{The $\nu$ exponent for  $(L,2L)$ pairs at the different
dilutions. The last column corresponds to an inverse linear extrapolation.}
\medskip
\begin{tabular}{rllll}\hline
\multicolumn{1}{c}{$p$}
    & \multicolumn{1}{c}{$L=8$}      
    & \multicolumn{1}{c}{$L=12$}      
    & \multicolumn{1}{c}{$L=16$}      
    & \multicolumn{1}{c}{$L=\infty$}\\ \hline
$0.8 $       &.5175(11)&.5154(11)&.5142(13)&.5110(25)\\
$0.65$       &.5308(13)&.5270(13)&.5251(12)&.5194(26)\\
$0.5 $       &.5482(16)&.5428(19)&.5412(19)&.534(4)\\
$\approx 0.4$&.5604(15)&.5532(19)&.5478(18)&.536(4)\\
$\approx 0.3$&.5700(26)&.5647(22)&.5583(26)&.549(5)\\\hline
\end{tabular}
\vglue -1 cm
\label{NUNAIVE}
\end{center}
\end{table}

We can check the expected logarithmic corrections directly
measuring $\xi$ and $C$ at the critical point.
To determine the critical coupling (or dilution),
we study the Binder $g_4$ parameter.

The hyperscaling violations for $\xi$ at the critical point are
\begin{equation}
\xi(L,\beta_{\rm c})\propto L \left(\log L\right)^{\delta_\xi}.
\end{equation}
In table \ref{XILOG} we show the fitted $\delta_\xi$ values, 
which are reasonably close to the predicted values.

\begin{table}[h]
\footnotesize
\begin{center}
\caption{$\delta_\xi$ for the different dilutions. The second error bar is
due to the uncertainty in the critical point and the first is the 
statistical one.}
\medskip
\begin{tabular}{ccccc}\cline{1-2}\cline{4-5}
      \multicolumn{1}{c}{$p$}      
    & \multicolumn{1}{c}{$\delta_\xi$}&
    & \multicolumn{1}{c}{$p$}      
    & \multicolumn{1}{c}{$\delta_\xi$}\\\cline{1-2}\cline{4-5}
1.0 &.198(3+5) &&0.8&.203(5+10)\\
0.65&.181(6+14)&&0.5&.189(7+12)\\
0.4 &.241(8+12)&&0.3&.281(10+23)\\\cline{1-2}\cline{4-5}
\end{tabular}
\label{XILOG}
\end{center}
\end{table}

\section{The $d=2$ model}
We perform $10^4$ samples for $L\leq 384$ at different dilutions.
PRG predicts logarithmic corrections to the Ising behavior: 
$\nu$, equal to one,  is corrected by $1/\log L$ 
and $C \propto \log (\log L)$. 
We find also good agreement to the predicted behavior in both cases.


\begin{thebibliography}{99}

\bibitem{HARRIS} A. B. Harris, {\sl J. Phys.} {\bf C7} (1974) 1671.
\bibitem{PARU} G. Parisi and J. J. Ruiz-Lorenzo,
{\sl J. Phys. A: Math. Gen.} {\bf 28} (1995) L395.
\bibitem{ISDIL4D} H.G. Ballesteros, L.A. Fern\'andez, V. Mart\'{\i}n-Mayor, 
A. Mu\~noz Sudupe, G. Parisi, J.J. Ruiz-Lorenzo,
{\sl hep-lat/9707017}.
\bibitem{ISDIL2D} 
H.G. Ballesteros, L.A. Fern\'andez, V. Mart\'{\i}n-Mayor, 
A. Mu\~noz Sudupe, G. Parisi, J.J. Ruiz-Lorenzo,
{\sl cond-mat/9707179}. To be published in J. Phys. {\bf A}.
\bibitem{PERC}
H.G. Ballesteros, L.A. Fern\'andez, V. Mart\'{\i}n-Mayor, 
A. Mu\~noz Sudupe, G. Parisi, J.J. Ruiz-Lorenzo,
{\sl Phys. Lett.} {\bf B400} (1997) 346. 


\end{thebibliography}
\end{document}